\documentclass[prd,showkeys,floatfix,nofootinbib, %twocolumn,
               preprint,12pt,tightenlines,fleqn]{revtex4-1}

\usepackage{amsmath,amssymb,revsymb,graphicx,dcolumn}
\usepackage{leftidx}
\newcommand  {\version}{v3}  %%true version v2.993 
\usepackage{slashed}

\usepackage{mathtools}

\newcommand{\beq}{\begin{equation}}
\newcommand{\eeq}{\end{equation}}
\newcommand{\beqa}{\begin{eqnarray}}
\newcommand{\eeqa}{\end{eqnarray}}
\newcommand{\bsubeqs}{\begin{subequations}}
\newcommand{\esubeqs}{\end{subequations}}

\usepackage{mwe}

\begin{document}
%\noindent CQG XX, yyyyy (2018)
%\hfill   arXiv:xxxxxxxxx%%%%\;(\version)
%
\noindent
arXiv:1807.08597
\hfill
KA--TP--16--2018\;(\version)
%
%arXiv:xxxxxxxxxx
%\preprint{KA--TP--16--2018\,(\today;\;\version)}
%
\newline\vspace*{3mm}

\title[]{Charge extraction from a black hole}

\author{V.A. Emelyanov}
\email{viacheslav.emelyanov@kit.edu}
\affiliation{Institute for
Theoretical Physics, Karlsruhe Institute of
Technology (KIT), 76128 Karlsruhe, Germany\\}

\author{F.R. Klinkhamer}
\email{frans.klinkhamer@kit.edu}
\affiliation{Institute for
Theoretical Physics, Karlsruhe Institute of
Technology (KIT), 76128 Karlsruhe, Germany\\}

\begin{abstract}
\vspace*{2.5mm}\noindent
Following up on earlier work about across-horizon quantum scattering
and information transfer outwards across the black-hole horizon,
we show that it is also possible to extract
electric charge from a static nonrotating black hole.
This result can be extended to other types of charge
with a gauge-boson coupling and to nonstatic rotating black holes.
\end{abstract}

\keywords{general relativity, black holes,
electroweak standard model, scattering}

\maketitle

\section{Introduction}
\label{sec:Introduction}

In a recent article~\cite{EmelyanovKlinkhamer2018}, we discussed
Coulomb scattering of two electrically charged
elementary particles, with one of these particles
inside the Schwarzschild black-hole horizon and the other outside.
We, then, proposed a \textit{Gedankenexperiment}
which uses this quantum scattering process to
transfer information from inside the black-hole horizon to outside.

Now, the question arises if it is, in principle, possible to extract
an electric charge from a static nonrotating black hole by the exchange
of a charged vector boson?
In the present note, we will give an affirmative answer to this question,
starting from an electron-positron pair
inside the Schwarzschild black-hole horizon.
This answer will be obtained by using the results from Ref.~\cite{EmelyanovKlinkhamer2018}
and those of an earlier preprint version~\cite{EmelyanovKlinkhamer2017v6}.

%%\newpage%%tmp
\section{Scattering set-up}
\label{sec:Scattering-set-up}

\begin{figure}[t]  %%figX-v091.eps=figX-v1.eps  %%fig1-v1.eps=fig1-v2.eps
\includegraphics[scale=0.5]{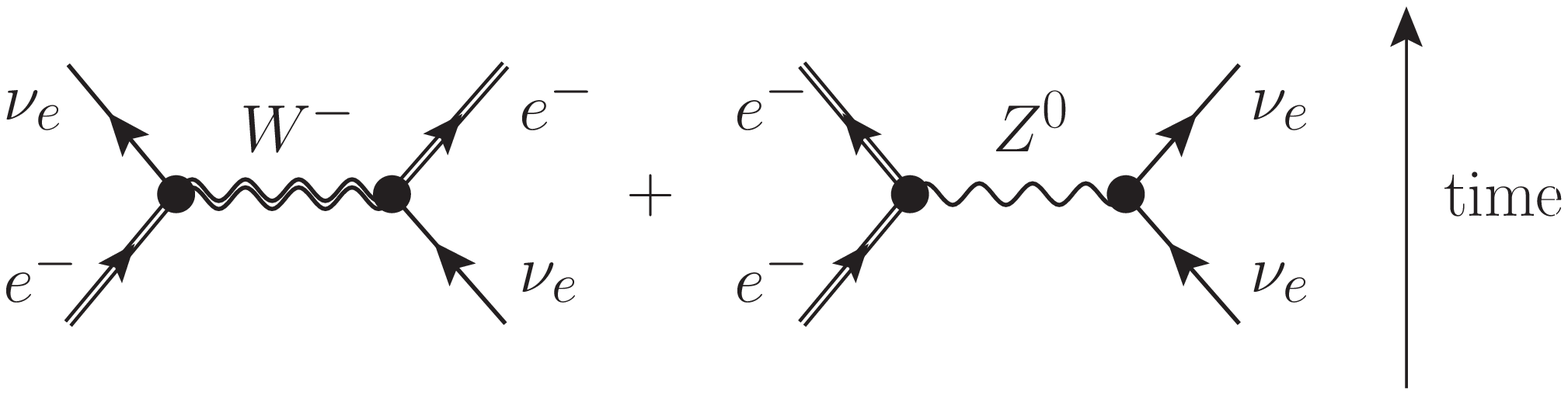}
\vspace*{-2mm}
\caption{Position-space Feynman diagrams for $e^{-}\,\nu_e$ scattering
at tree level in Minkowski spacetime.
The arrows in the diagrams show the flow of lepton number.
The double lines indicate that electric charge is transported.
}
\label{fig:1}
%\end{figure}
\vspace*{10mm}
%\begin{figure}[h]
\includegraphics[scale=0.5]{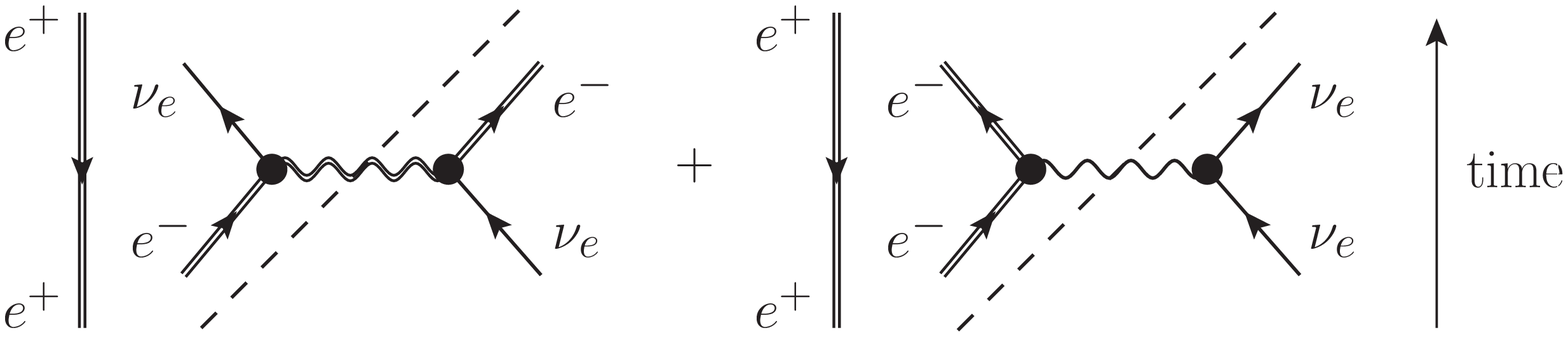}
%{charge-extraction-fig2-v1031.eps}
\vspace*{-2mm}
\caption{Across-horizon $e^{-}\,\nu_e$ scattering
allows for electric-charge extraction from a static nonrotating black hole.
The position-space Feynman diagrams from Fig.~\ref{fig:1}
now hold in a local inertial coordinate system
near the Schwarz\-schild black-hole horizon.
Specifically, the electric-charge-extraction process follows from
the charged-current position-space Feynman diagram on the left.
The projected black-hole horizon is indicated
by the dashed line with the black-hole center to its left.
The positron inside the black-hole horizon does not participate
in the scattering.}
\label{fig:2}
\end{figure}

Consider the following elastic scattering process
from the standard model of elementary particles in Minkowski spacetime:
\beq
\label{eq:e-nue-scattering}
e^{-} + \nu_e \rightarrow e^{-} + \nu_e\,,
\eeq
as discussed in, e.g., Sec.~8.5 of Ref.~\cite{Taylor1976}
and Sec.~12.6.2 of Ref.~\cite{ItzyksonZuber1980}.
The relevant position-space Feynman diagrams at tree level
are given in Fig.~\ref{fig:1}.

The corresponding process in the black-hole context
is given by the set-up of Fig.~\ref{fig:2},
which needs to be compared to Figs.~2 and 3 of Ref.~\cite{EmelyanovKlinkhamer2018}.
For the set-up of Fig.~\ref{fig:2} in this note,
we also assume that the initial inside-horizon electron ($e^{-}$)
was produced by pair creation and
that the corresponding initial inside-horizon positron ($e^{+}$)
does not participate in the scattering with the
initial outside-horizon neutrino ($\nu_e$).
The total electric charge of this
initial inside-horizon positron-electron pair is zero.
\vspace*{0mm}

%%\newpage%%tmp
\section{Charge extraction}
\label{sec:Charge-extraction}

The position-space Feynman diagram on the left of
Fig.~\ref{fig:2} allows for electric charge extraction, as long as
the initial electron is inside the event horizon and
the final electron outside the event horizon
with an appropriate outgoing momentum
(similar to the recoil electron in elastic $e^{-}\,\mu^{-}$ scattering
as discussed in Ref.~\cite{EmelyanovKlinkhamer2018}).
Observe that the electric-charge-extraction process from Fig.~\ref{fig:2}
does not change the total lepton number inside the black-hole horizon.

The Einstein Equivalence Principle allows us to analytically describe this
process by employing the Minkowski-spacetime approximation for the
computation of $S$-matrix. 
An essential part of this approximation is
that we can neglect spacetime curvature effects contributing to
the diagrams of Fig.~\ref{fig:2} by considering initial particles with a
sufficiently large center-of-mass energy~\cite{Endnote1}.
In other words, the characteristic
length scale describing the scattering reaction has to be much smaller
than the curvature length scale near the event horizon for the
approximation to be reliable. 
This condition can always be fulfilled
by making the black-hole mass in the            
\textit{Gedankenexperiment} arbitrarily large~\cite{Endnote2}. 
Further discussion
of the near-horizon region  of a Schwarzschild black hole
and the local Minkowski coordinates $(T,\, X,\, Y,\, Z)$
can be found in Sec.~2 of Ref.~\cite{EmelyanovKlinkhamer2018}.

From now on, we will use the particle-physics conventions
of Sec.~20.2 in Ref.~\cite{PeskinSchroeder1995}.
The tree-level \mbox{$W$-exchange} diagram of Fig.~\ref{fig:2}
corresponds to the following momentum-space probability amplitude:
\beqa
\label{eq:mom-space-amplitude}
\mathcal{M}_{W}(e^{-}\,\nu_{e} \rightarrow e^{-}\,\nu_{e}) &=&
- \frac{g^2}{2}\,\frac{1}{q^2 -m_{W}^2 + i\varepsilon}\,
\left[\bar{u}^{s'}(k')\gamma_\mu u^r(p)\right]
\;\left[\bar{u}^{r'}(p')\gamma^\mu u^s(k)\right]\,,
\eeqa
with the $SU(2)$ gauge coupling constant $g$
and the gauge-boson 4-momentum
$q \equiv k-p' = k'-p$ for 4-momenta
$p/p'$ and $k/k'$ of the initial/final neutrino and electron,
respectively. On the left-hand side of \eqref{eq:mom-space-amplitude},
we have suppressed the spin indices
$r,\,r'$ of the neutrino and $s,\,s'$ of the electron.

Next, specialize to spin-up fermions, zero neutrino mass, and
the following 4-momenta:%
\bsubeqs
\beqa
\label{eq:choice-momenta}
k^{\,\mu} &=& \left(\sqrt{k^2 + m_{e}^2},\,k,\,0,\,0 \right)\,,
\\[2mm]
p^{\,\mu} &=& \left(p,\,-\sqrt{1/3}\;p,\,+\sqrt{2/3}\;p,\,0 \right)\,,
\\[2mm]
k'^{\,\mu} &=& k^{\,\mu} \,,
\\[2mm]
p'^{\,\mu} &=& p^{\,\mu} \,,
\eeqa
\esubeqs
where both $k$ and $p$ are taken positive. Then, we have
\beqa\hspace{-10mm}
\label{eq:mom-space-amplitude-result}
\mathcal{M}_{W}(e^{-}\,\nu_{e} \rightarrow e^{-}\,\nu_{e}) &=&
\frac{2\,g^2}{\sqrt{3}}\;
\frac{p\,k}{m_{e}^2 - \sqrt{4/3}\;p\,k - 2\,p\,\sqrt{k^2 + m_{e}^2} -m_{W}^2
            + i\varepsilon}
\nonumber\\[1mm]
&\sim&
\frac{g^2}{2}\,\left(1-\sqrt{3}\right) \;\neq\; 0\,,
\eeqa
in the ultrarelativistic limit of $k$ and $p$
[specifically, $\min(k,\,p) \gg m_{W} \gg m_{e}$].
For the calculation of the position-space
amplitude shown on the left of Fig.~\ref{fig:2},  we must
fold the above momentum-space amplitude with appropriate
wave packages, as discussed in Appendix C of an earlier
version of our article~\cite{EmelyanovKlinkhamer2017v6}.

The nonvanishing $e^{-}\,\nu_e$ scattering amplitude
\eqref{eq:mom-space-amplitude-result} implies,  according to
the argument of Secs.~2 and 3 in Ref.~\cite{EmelyanovKlinkhamer2018}
and App. C in Ref.~\cite{EmelyanovKlinkhamer2017v6},
that we can extract electric charge from a static nonrotating
black hole by repeating the scattering process of Fig.~\ref{fig:2}
a large number of times ($N \gg 1$). Only a few events ($n \ll N$)
extract electric charge, as both Feynman diagrams of Fig.~\ref{fig:2}
contribute to the process and
even the charged-current diagram may not give an
appropriate outgoing electron
(cf. App.~B in Ref.~\cite{EmelyanovKlinkhamer2018}).

%%\newpage%%tmp
\section{Discussion}
\label{sec:Discussion}

\begin{figure}[t]
\includegraphics[scale=0.55]{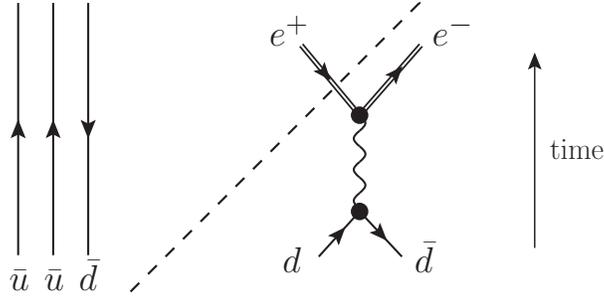}
%{charge-extraction-fig3-v1031.eps}
%{charge-extraction-fig3-v09989.eps}
\vspace*{-2mm}
\caption{Position-space Feynman diagram
(in a near-horizon local inertial coordinate system)
for electric-charge reduction of a static nonrotating charged black hole.
Shown is a black hole with an initial negative electric charge $Q=-e$,
which is changed to $Q=0$ by an infalling positron
(electric charge $e>0$  and lepton number $-1$),
while the corresponding electron
(electric charge $-e<0$ and lepton number $1$) escapes to spatial infinity.
The double line denotes an electron and the single line a quark
(up or down quark with electric charge $2e/3$ or $-e/3$, respectively,
each quark having baryon number $1/3$),
where the arrows indicate the flow of negative electric charge.
The exchange particle in the sub-diagram on the right can be  %%FRK
either a photon or a $Z^{0}$ boson.                           %%FRK   
The projected black-hole horizon is shown
by the dashed line with the black-hole center to its left.}
\label{fig:3}
\end{figure}

It is, of course, possible to change the electric charge
of a static nonrotating black hole by other processes than the
one considered up till now.
Different from the electric-charge-extraction process
(left-diagram of Fig.~\ref{fig:2}) is,
for example, the electric-charge-reduction process as illustrated by Fig.~\ref{fig:3}.
Observe that the electric-charge-reduction process from Fig.~\ref{fig:3}
changes the total lepton number inside the black-hole horizon.
Note also that, for both processes,
the backreaction on the metric has been neglected
in Figs.~\ref{fig:2} and \ref{fig:3}.

Returning to our electric-charge-extraction process (Fig.~\ref{fig:2}),
several remarks are in order.
First, the very process of charge extraction considered
in Sec.~\ref{sec:Charge-extraction}
does not imply that causality is violated, as there is no real
(on-shell) particle that moves superluminally.

Second, we have thus seen that a quantum scattering process
not only allows for the extraction of information from a
static nonrotating black hole
but also for the extraction of electric charge.
In this respect, we note that
electric charge is not just a measure of how strongly
two electrons scatter with each other, but, rather,
a measure of how the corresponding photon propagates
(cf. the discussion on electric charge in Sec.~III.7,
pp. 204--205 of Ref.~\cite{Zee2010}).
This becomes clear if we examine the structure
of radiative corrections contributing to charge
renormalization. It turns out that charge
renormalization depends only on the photon
propagator (including vacuum-polarization effects).
Thus, we can say that the electric charge belongs
to the gauge field
(determining how the photon propagates),
rather than to the matter field.

Third, our result for electric-charge extraction
from a static nonrotating black hole can be extended
to other types of charge involving a gauge boson
and to nonstatic rotating black holes.

%%\newpage%%tmp
\vspace*{-0mm}
\section*{\hspace*{-5mm}Acknowledgments}
\vspace*{-0mm}\noindent
We thank P. Soler for asking,
after an invited talk by V.A.E. at the May 2018 Heidelberg meeting of
TRR33 ``The Dark Universe,''
the question formulated in Sec.~\ref{sec:Introduction} of the present note.


\newpage
\begin{thebibliography}{9}

\bibitem{EmelyanovKlinkhamer2018}
V.A.~Emelyanov and F.R.~Klinkhamer,
``Across-horizon scattering and information transfer,''
Class.\ Quant.\ Grav.\  {\bf 35},  125004 (2018)
[arXiv:1710.06405].
%%CITATION = doi:10.1088/1361-6382/aac1f1;%%

\bibitem{EmelyanovKlinkhamer2017v6}
V.A.~Emelyanov and F.R.~Klinkhamer,
``Across-horizon scattering and information transfer,''
arXiv:1710.06405v6.


\bibitem{Taylor1976}
J.C.~Taylor,
\emph{Gauge Theories of Weak Interactions}
(Cambridge Univ. Press, Cambridge, England, 1976).
  %%CITATION = INSPIRE-113517;%%

\bibitem{ItzyksonZuber1980}
C.~Itzykson and J.B.~Zuber,
\textit{Quantum Field Theory}
(McGraw--Hill, New York, USA, 1980).

\bibitem{Endnote1}
Note that the quantum vacuum 
in the observable Universe \emph{locally} provides
a Minkowski-vacuum representation for particle physics. 
Hence, it makes, \emph{a posteriori},
physical sense to compare the outcomes of scattering experiments 
at CERN with those at SLAC.

\bibitem{Endnote2}
Note that the asymptotic states in the $S$-matrix definition 
(states from the limits $t \rightarrow \pm\infty$)
are a mathematical abstraction. 
Physically, these states correspond to well-separated wave packages, 
each of which can be considered to describe a free elementary particle. This is consistent with actual measurements in collider-physics experiments, where initial and final ``asymptotic'' states are
typically separated by a tiny fraction of a second.

\bibitem{PeskinSchroeder1995}
M.E. Peskin and D.V. Schroeder,
\emph{An Introduction to Quantum Field Theory}
(Addison-Wesley Publishing Co., Reading, MA, USA, 1995).
%%CITATION = INSPIRE-407703;%%

\bibitem{Zee2010}
A.~Zee,
\emph{Quantum Field Theory in a Nutshell}, Second Edition
(Princeton Univ. Press, Princeton, NJ, USA, 2010).

\end{thebibliography}
\end{document}